\documentclass[aps,prb,amsmath,amssymb,twocolumn,superscriptaddress]{revtex4-1}
\usepackage{amsmath,graphicx,color,verbatim}
\usepackage{graphicx}
\usepackage{xspace}
\usepackage{longtable}
\usepackage{booktabs}
\def\beq{\begin{equation}}
\def\eeq{\end{equation}}

\newcommand{\rrscan}{r\textsuperscript{2}SCAN\xspace}

\begin{document}

\title{Testing the r$^2$SCAN  density functional for the
thermodynamic stability of solids with and without a van der
Waals correction}

\author{Manish Kothakonda}
\affiliation{Department of Physics and Engineering Physics, Tulane University, New Orleans, Louisiana 70118, United States}

\author{Aaron D. Kaplan}
\affiliation{Department of Physics, Temple University,
Philadelphia, Pennsylvania 19122, United States}
\author{Eric B. Isaacs}
\affiliation{HRL Laboratories, LLC, Malibu, California 90265, United States}
\author{Christopher J. Bartel}
\affiliation{ Department of Chemical Engineering and Materials Science, University of Minnesota, Minneapolis, Minnesota 55455, United States}
\author{James W. Furness}
\affiliation{Department of Physics and Engineering Physics, Tulane University, New Orleans, Louisiana 70118, United States}
\author{Jinliang Ning}
\affiliation{Department of Physics and Engineering Physics, Tulane University, New Orleans, Louisiana 70118, United States}

\author{Chris Wolverton}
\affiliation{Department of Materials Science and Engineering, Northwestern University, Evanston, Illinois 60208, United States}

\author{John P. Perdew}
\affiliation{Department of Physics, Temple University,
Philadelphia, Pennsylvania 19122, United States}
\author{Jianwei Sun}
\email{jsun@tulane.edu}
\affiliation{Department of Physics and Engineering Physics, Tulane University, New Orleans, Louisiana 70118, United States}

\begin{abstract}

A central aim of materials discovery is an accurate and numerically reliable description of thermodynamic properties, such as the enthalpies of formation and decomposition.
The \rrscan revision of the strongly constrained and appropriately normed (SCAN) meta-generalized gradient approximation (meta-GGA) balances numerical stability with high general accuracy.
To assess the \rrscan description of solid-state thermodynamics, we evaluate the formation and decomposition enthalpies, equilibrium volumes, and fundamental bandgaps of more than 1,000 solids using \rrscan, SCAN, and PBE, as well as two dispersion-corrected variants, SCAN+rVV10 and \rrscan{}+rVV10.
We show that \rrscan{} achieves  accuracy comparable to SCAN and often improves upon SCAN's already excellent accuracy.
Whereas SCAN+rVV10 is often observed to worsen the formation enthalpies of SCAN, and makes no substantial correction to SCAN's cell volume predictions, \rrscan{}+rVV10 predicts marginally less-accurate formation enthalpies than \rrscan{}, and slightly more-accurate cell volumes than \rrscan{}.
The average absolute errors in predicted formation enthalpies are found to
decrease by a factor of 1.5 to 2.5 from the GGA level to the meta-GGA level.
Smaller decreases in error are observed for decomposition enthalpies.
For formation enthalpies \rrscan improves over SCAN for intermetallic systems. 
For a few classes of systems -- transition metals, intermetallics, weakly-bound solids, and enthalpies of decomposition into compounds -- GGAs are comparable to meta-GGAs.
In total, \rrscan{} and \rrscan{}+rVV10 can be recommended as stable, general-purpose meta-GGAs for materials discovery.

\end{abstract}

\date{\today}
\maketitle

\section{Introduction}

The backbone of modern \textit{ab initio} simulations of solids is practical Kohn-Sham density functional theory (DFT) \cite{kohn1965}. Efficient, first principles approximations to the generally-unknown exchange-correlation energy have made rapid advances in solid state materials physics possible. Within the Perdew-Schmidt \cite{perdew2001} hierarchy of density functional approximations (DFAs), the generalized gradient approximation (GGA), which depends upon the spin-densities and their gradients, and the meta-GGA, which further depends on the local kinetic energy spin-densities, stand as the most appealing semi-local DFAs.
GGAs, like the Perdew-Burke-Ernzerhof (PBE) GGA \cite{perdew1996}, offer reasonable general accuracy at low computational expense.

Meta-GGAs offer greater general accuracy than GGAs, but can be much more computationally intensive to use.
The strongly constrained and appropriately normed (SCAN) meta-GGA \cite{sun2015} accurately simulates complex materials, such as the ``strongly-correlated'' cuprates \cite{Furness2018,Zhang2020b} and transition metal monoxides \cite{Zhang2020c}, but suffers well-known numerical instabilities inherent to its construction \cite{Bartok2019,Furness2019}.
These instabilities are more problematic for small-grid codes used to study atoms, molecules, and clusters, but can also make stable convergence behavior challenging in plane-wave codes used for solids \cite{ning2022}.

The r$^2$SCAN meta-GGA \cite{furness2020} was constructed as a numerically-stable, general-purpose revision of SCAN intended to retain much of its accuracy. r$^2$SCAN builds upon the rSCAN meta-GGA of Bart\'ok and Yates \cite{Bartok2019}, but restores important exact constraints to rSCAN, such as the uniform density limit and coordinate scaling properties \cite{furness2022}. Tests of \rrscan for molecules \cite{furness2020,Mejia-Rodriguez2020,Mejia-Rodriguez2020g,ehlert2021,Grimme2021} and for solids \cite{furness2020,Mejia-Rodriguez2020,kingsbury2022,furness2022,ning2022reliable} have shown that \rrscan indeed retains or improves upon the high accuracy of SCAN.

\begin{figure}
    \centering
    {\includegraphics[width=0.48\textwidth]{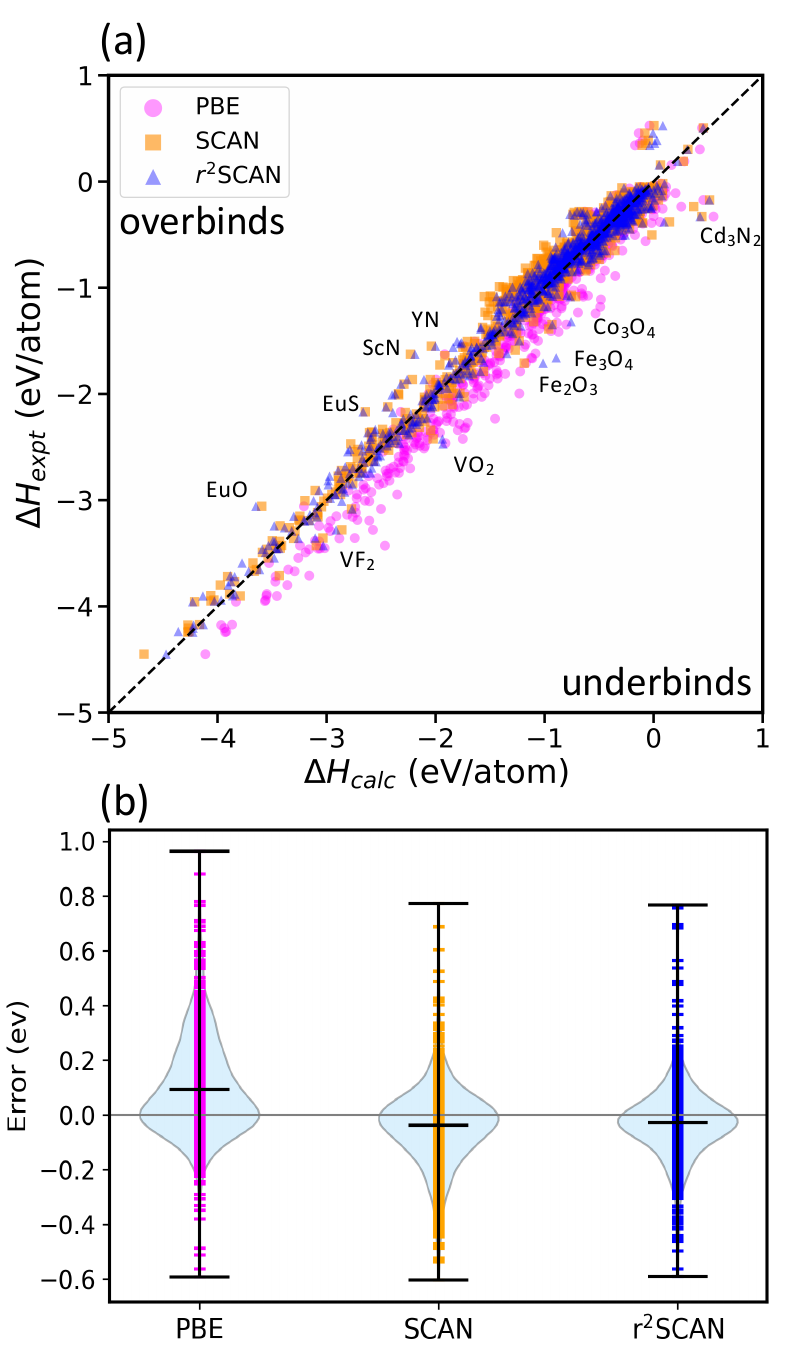}}
    \caption{(a) Comparison of calculated and experimental formation energy for the 1015 compounds for PBE, SCAN and r$^2$SCAN. The dashed diagonal line corresponds to the $\Delta H_\mathrm{calc}$ = $\Delta H_\mathrm{expt}$ line of perfect agreement. (b) Violin plots of the error distributions in the solid set. The \rrscan{} median error lies closest to zero.}
    \label{fig:formation_energy}
\end{figure}

To promote further progress towards high-throughput meta-GGA calculations for solids, we compare the formation and decomposition enthalpies, unit cell volumes, and electronic structures of more than 1,000 solid-state materials calculated using \rrscan{}, SCAN, and PBE. 
In addition to SCAN and \rrscan, we present results for their dispersion-corrected \cite{sabatini2013} variants: SCAN+rVV10 \cite{peng2016} and \rrscan{}+rVV10 \cite{ning2022}.
The dispersion-corrected \rrscan{}+D4 \cite{ehlert2021} describes molecular thermochemistry with exceptional accuracy, however a broad benchmark of a dispersion-corrected \rrscan in solids has not previously been attempted.
As in Refs. \citenum{peng2016} and \citenum{ning2022}, we use $b=15.7$ for SCAN+rVV10 and $b=11.95$ for \rrscan{}+rVV10.
The $b$-parameter controls the damping of the rVV10 dispersion correction at short-range.
A larger $b$ produces a stronger cutoff.
This is needed as semilocal DFAs include a reasonable description of short-range correlation, and meta-GGAs in particular can include an accurate description of intermediate-ranged dispersion interactions in their exchange parts.

\section{Computational details}

Calculation of the enthalpies of formation are performed for 934 binary compounds and 81 ternary compounds (see Table S1 of the Supplementary Materials for chemical formulas). The structures and reference formation enthalpies for these 1015 compounds are taken from the datasets of Isaacs \textit{et al.} \cite{isaacs2018} and Zhang \textit{et al.} \cite{zhang2018}. Reference structures and enthalpies of decomposition for 987 compounds are taken from the dataset of Bartel \textit{et al.}\cite{bartel2019role}.

All calculations are performed using the Vienna Ab Initio Simulation Package (VASP) \cite{Kresse1993,Kresse1994,Kresse1996,Kresse1996a} using the projector augmented wave (PAW) method. A plane wave energy cutoff of 600 eV is used. $\Gamma$-centered, uniform Monkhorst-Pack $k$-point meshes with $k$-point density of 700 $k$-points per \AA{}$^{-3}$ are generated with pymatgen \cite{jain2011high}. First-order Methfessel-Paxton smearing \cite{methfessel1989} of width 0.2 eV is employed for structural relaxations, while total energy calculations use the tetrahedron method with Bl\"ochl corrections \cite{blochl1994}. We compare three semilocal exchange-correlation density functional approximations (DFAs): the PBE GGA\cite{perdew1996}, the SCAN meta-GGA \cite{sun2015}, and the r$^2$SCAN meta-GGA \cite{furness2020}. As no meta-GGA pseudopotentials are available in VASP, we use the ``PAW 52'' PBE pseudopotentials. In magnetically active systems, the ferromagnetic ordering is considered to be the ground-state. For the systems CrB, CoF$_2$, CNiO$_3$, F$_2$Mn, Fe$_2$O$_3$, Fe$_3$O$_4$,  Fe$_4$Ni$_2$O$_8$, and NiSO$_4$ antiferromagnetic orderings are considered. For structure selection, the calculations are converged to $10^{-6}$ eV in the total energy, and 0.01 eV/\AA{} in the atomic forces. For computing formation enthalpies, all calculations are converged to $10^{-7}$ eV in the total energy, and 0.01 eV/\AA{} in the atomic forces. Molecular reference states are used for H$_2$, N$_2$, O$_2$, F$_2$ and Cl$_2$, where the isolated molecule is represented by a dimer in a $15 \times 15 \times 15$ \AA{}$^3$ box. Experimental standard enthalpies of formation used to determine the error in formation energy are defined at 298 K and 1 atm of pressure \cite{isaacs2018}.

\section{Results and Discussion}

\subsection{Formation enthalpy}

\begin{figure*}
    \centering
    {\includegraphics[width=1.0\textwidth]{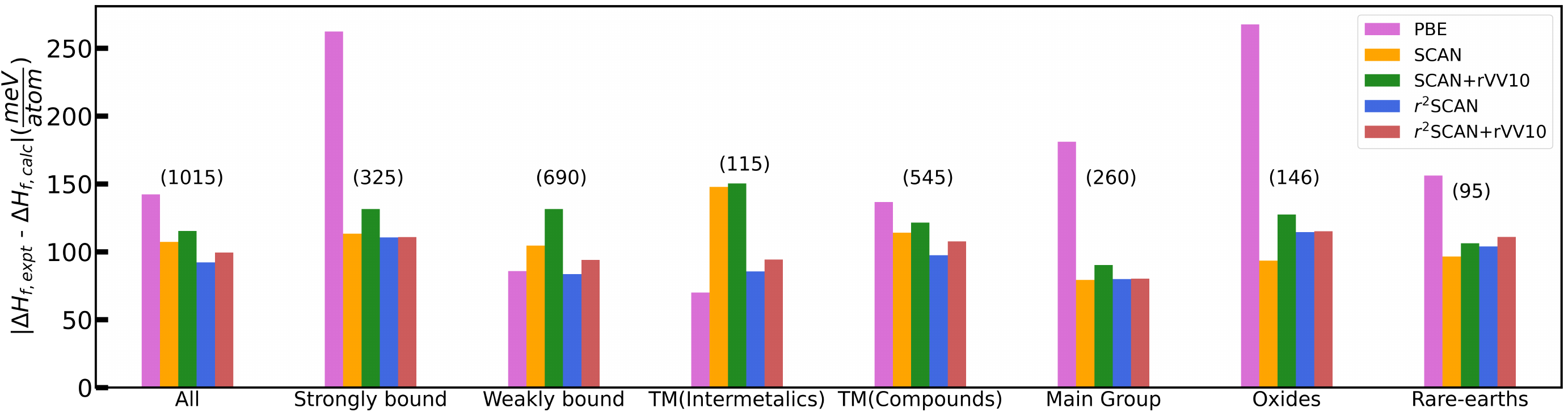}}
    \caption{Comparison of mean absolute errors for PBE, SCAN, SCAN+rVV10, r$^2$SCAN, and r$^2$SCAN+rVV10 with respect to experimental values for formation enthalpies of solids.
    The 1015 set is partitioned into subsets defined in the text. The numbers in parentheses above each set of bars indicate the number of compounds in that subset. }
    \label{fig:MAD_fe}
\end{figure*}

To systematically compare the performance of DFAs on formation enthalpies of solid-state materials, we group databases according to Isaacs \textit{et al.} \cite{isaacs2018} and Zhang \textit{et al.} \cite{zhang2018}. The total set comprises 1015 solids. Figure \ref{fig:formation_energy}(a), which compares experimental and calculated formation enthalpies, shows that PBE systematically underbinds solids, whereas SCAN and \rrscan{} tend to overbind ``weakly-bound'' solids ($|\Delta H_\text{expt}| \lesssim 1$ eV/atom).
Violin plots of the PBE, SCAN, and \rrscan{} error distributions are shown in Fig. \ref{fig:formation_energy}(b).
The PBE error distribution is strongly skewed towards positive errors (predicting too-small absolute formation enthalpies), indicating systematic underbinding.
SCAN's errors are much less systematic, but show a tendency to slightly overbind. 
The \rrscan{} median error lies closest to zero, and the error distribution is more symmetric than SCAN's.

While we have performed calculations using the rVV10 counterparts of SCAN and \rrscan, they are not presented in Fig. \ref{fig:formation_energy} for reasons of clarity.
Scatter plots of the errors made by the rVV10-corrected meta-GGAs are given in Fig. \ref{fig:sm_fe} of the Supplementary Materials.

To better gauge the accuracy of predicted formation enthalpies, Fig. \ref{fig:MAD_fe} presents errors for subsets of the database.
``All'' is the entire 1015 solid set; ``strongly-bound'' solids have experimental formation enthalpies $-4 \leq \Delta H_\text{expt} \leq -1$ eV/atom; ``weakly-bound'' solids have $|\Delta H_\text{expt}| < 1$ eV/atom.
Transition metal (TM)-containing compounds are grouped into TM(Intermetallics), which are intermetallics composed only of transition metals, and TM(Compounds), which contain other elements.
Main group solids contain elements from the main group (groups 1, 2, and 13--18 of the periodic table).
Oxides are oxygen-containing solids, and rare-earths contain at least one rare-earth element (lanthanide series, Sc and Y).

We define a few statistical error metrics that will be used throughout: the mean error (ME) or mean deviation (MD)
\begin{equation}
    \text{ME/MD} = \frac{1}{N} \sum_{i=1}^N (X_i^\text{DFA} - X_i^\text{ref}),
\end{equation}
where $X_i^\text{DFA}$ is a quantity (energy difference, volume, bandgap, etc.) computed with a DFA, and $X_i^\text{ref}$ is a reference value.
We assume $N$ quantities belong to a set.
We use ``error'' to indicate that a reference value is known with very low uncertainty and high accuracy. We use ``deviation'' when comparing quantities between different approximate methods.
The mean absolute error (MAE) or deviation (MAD) is
\begin{equation}
    \text{MAE/MAD} = \frac{1}{N} \sum_{i=1}^N |X_i^\text{DFA} - X_i^\text{ref}|.
\end{equation}
When analyzed in conjunction with the MAE/MAD, the ME/MD is useful for determining the degree to which a DFA makes systematic errors.
If $|\text{ME}|=\text{MAE}$, a DFA makes wholly systematic errors.
If $|\text{ME}|\approx 0$, a DFA makes essentially random errors.
The root-mean-squared error (RMSE) or deviation (RMSD)
\begin{equation}
    \text{RMSE/RMSD} = \left[\frac{1}{N} \sum_{i=1}^N (X_i^\text{DFA} - X_i^\text{ref})^2 \right]^{1/2},
\end{equation}
is a metric comparable to the MAE/MAD.
The RMSE/RMSD is simply the square root of the variance.
The MAE/MAD is more frequently used than the RMSE/RMSD, however both carry important information.

For the entire set, the 92 meV/atom MAE of \rrscan is the lowest of all considered DFAs, including \rrscan{}+rVV10 (99 meV/atom MAE).
SCAN has a modestly higher 107 meV/atom MAE for the entire set.
For strongly-bound compounds, \rrscan, \rrscan{}+rVV10, and SCAN have nearly identical $\sim$111 meV/atom MAEs. 
\rrscan{} and PBE predict the most accurate formation enthalpies for weakly-bound solids with 84 and 86 meV/atom MAEs, respectively.
SCAN and SCAN+rVV10 find larger errors for these solids with 105 and 132 meV/atom MAEs, respectively.
Consistent with Refs. \citenum{isaacs2018} and \citenum{kingsbury2022}, PBE predicts the most accurate formation enthalpies of intermetallics (70 meV/atom MAE), with SCAN and SCAN+rVV10 making substantially larger MAEs, 148 and 150 meV/atom, respectively.
\rrscan{} and \rrscan{}+rVV10 predict intermetallic formation enthalpies with accuracy much closer to PBE: their MAEs are 86 and 94 meV/atom, respectively.

For transition metal-containing compounds, \rrscan{} has the lowest MAE at 97 meV/atom, followed by \rrscan{}+rVV10, SCAN, SCAN+rVV10 and PBE, with 108, 114, 122, 137 meV/atom MAEs, respectively.
For main group compounds, \rrscan, SCAN, and \rrscan{}+rVV10 have nearly identical $\sim$80 meV/atom MAEs.
SCAN is the most accurate DFA for the oxides (94 meV/atom), with \rrscan{} and \rrscan{}+rVV10 following closely behind (114 and 115 meV/atom MAEs, respectively).
Last, all meta-GGAs are comparably accurate for the rare-earth-containing compounds, with SCAN making the smallest MAE, 97 meV/atom.

SCAN+rVV10 often predicts markedly less accurate formation enthalpies than SCAN, as is the case for the strongly-bound, weakly-bound, and oxide compounds of Fig. \ref{fig:MAD_fe}.
The increase in errors made by \rrscan{}+rVV10 over \rrscan{} is generally less pronounced.
SCAN already includes a large fraction of intermediate-range dispersion interactions in its exchange functional, indicated by the large $b$-damping parameter.
Thus, SCAN+rVV10 often further overbinds solids that SCAN overbinds.
\rrscan{} includes a less comprehensive description of intermediate-range dispersion interactions than SCAN (indicated by the smaller $b$ value, or less severe damping).
Thus, even in cases where \rrscan{} overbinds, \rrscan{}+rVV10 further overbinds, but to a less pronounced extent.

\subsection{Volumes}

To assess the accuracy of the predicted crystal structures, we compare the computed relaxed volume per atom to experimental values. Table \ref{tab:veq_errs} presents errors in the equilibrium volumes predicted by PBE, SCAN, r$^2$SCAN, and their rVV10 counterparts. While PBE overestimates volumes by 0.77 \AA{}$^3$/atom on average, SCAN underestimates equilibrium volumes by 0.11 \AA{}$^3$/atom on average, and r$^2$SCAN overestimates them by 0.24 \AA{}$^3$/atom. The MAE in equilibrium volumes for r$^2$SCAN and SCAN are 0.59 \AA{}$^3$/atom and 0.58 \AA{}$^3$/atom respectively; thus r$^2$SCAN retains the good general accuracy of SCAN. 

The rVV10 van der Waals (vdW) correction does not improve upon the volumes predicted by SCAN.
However, \rrscan{}+rVV10 improves slightly on \rrscan with a 0.5 \AA{}$^3$/atom MAE.
This is again a reflection of the underlying meta-GGA description of dispersion interactions.
rVV10 often produces more meaningful corrections to \rrscan{} than to SCAN because SCAN includes a more substantial description of intermediate-range dispersion interactions.
Thus rVV10 can often over-correct SCAN.

\begin{table}[h]
    \centering
    \caption{Statistical errors in equilibrium volumes (\AA{}$^3$/atom) for a few density functional approximations (DFAs): PBE, SCAN, SCAN+rVV10, \rrscan, and \rrscan{}+rVV10.
    \rrscan preserves much of the accuracy of SCAN at better computational efficiency.
    SCAN+rVV10 performs as accurately as SCAN, with a tendency to predict slightly smaller volumes.
    \rrscan{}+rVV10 offers a slight improvement over \rrscan.
    }
    \begin{tabular}{lrrrrr} \hline\hline
        DFA & ME && MAE && RMSE \\
        & (\AA{}$^3$/atom) && (\AA{}$^3$/atom) && (\AA{}$^3$/atom) \\\hline\hline
        PBE & 0.77 && 0.98 && 1.80 \\
        SCAN & -0.11 && 0.58 && 0.96 \\
        SCAN+rVV10 &-0.32 && 0.59 &&0.95 \\
        \rrscan{} & 0.24 && 0.59 && 1.04  \\
        \rrscan{}+rVV10 & -0.11 && 0.5 && 0.88 \\ \hline
    \end{tabular}
    \label{tab:veq_errs}
\end{table}

Notably, \rrscan{} and SCAN over- and underestimate the volume of CoI$_2$ by 7\% (2.2 \AA{}$^3$/atom and -2.1\AA{}$^3$/atom), respectively; \rrscan{}+rVV10 overestimates its volume by only 1.6\% (0.5 \AA{}$^3$/atom).
The volumes of layered materials tend to be more accurate when a vdW correction is used\cite{ning2022,peng2016}.
Thus, using a vdW correction to \rrscan{} or SCAN can improve cell volumes without harming the accuracy of predicted formation enthalpies, and can be recommended for general materials discovery.

\subsection{Magnetism}

Next, we explore the magnetic properties of the elemental metals Fe, Co, and Ni using PBE, SCAN and r$^2$SCAN. The predicted and experimental saturation magnetizations are shown in Table \ref{tab:magnetism}. In all cases, r$^2$SCAN predicts larger magnetic moments than SCAN, which in turn predicts larger magnetic moments than PBE. r$^2$SCAN and SCAN overestimate the magnetization of Fe by 24\% and 17\% respectively, while PBE underestimates it by only 1.8\%. In contrast, SCAN's magnetization for Co (1.72 $\mu_B$) is closer to the experimental value (1.75 $\mu_B$) than that of r$^2$SCAN (1.78 $\mu_B$) and PBE (1.59 $\mu_B$). These results confirm a known tendency \cite{mejia2019analysis} of r$^2$SCAN to overestimate magnetic moments. 

\begin{table}[h]
\centering
\caption{Magnetic moments of Fe, Co, and Ni computed using PBE, SCAN, SCAN+rVV10, \rrscan, and \rrscan{}+rVV10. Experimental values are included for comparison.}
\label{tab:magnetism}
\begin{tabular}{lrrrrrrr}

\hline\hline
    DFA         &  & & Fe($\mu_B$)   &  & Co($\mu_B$)   &  & Ni($\mu_B$)   \\ \hline\hline
PBE          &  &  &2.18 &  & 1.59 &  & 0.62 \\
SCAN         &  &  &2.60 &  & 1.72 &  & 0.72 \\
SCAN+rVV10   &  & & 2.66 &  & 1.77 &  & 0.82 \\
\rrscan{}    &  & &2.76 &  & 1.78 &  & 0.80 \\
\rrscan{} +rVV10 &  & &2.75 &  & 1.78 &  & 0.79 \\
Experiment   &  & &2.22 &  & 1.75 &  & 0.62 \\ \hline
\end{tabular}

\end{table}

While SCAN+rVV10 predicts larger magnetic moments than SCAN, \rrscan{}+rVV10 predicts nearly the same magnetic moments as \rrscan{}.
The local magnetic moments predicted by PBE, SCAN (+rVV10), and \rrscan{} (+rVV10) for all magnetic systems with magnetic moment greater than 0.1$\mu_B$ are shown in Supplementary Materials Fig. \ref{fig:avgmagentimoment}. 
\rrscan and SCAN predict 15\% and 12\% larger magnetic moments (on average) than PBE; their rVV10 counterparts show slightly lower average magnetic moments in comparison with the meta-GGAs.

\subsection{Bandgaps}

Here we consider SCAN, r$^2$SCAN, and their rVV10 counterparts for electronic bandgap prediction. It is well known that semilocal DFAs such as PBE underestimate the fundamental bandgap\cite{perdew1985density}. 
In a GGA or a meta-GGA (when the latter is implemented in a generalized Kohn-Sham scheme), the fundamental bandgap for a given DFA equals the ionization energy minus the electron affinity of the solid for the same DFA.
The meta-GGA bandgaps tend to be slightly more realistic than those of GGAs because the corresponding total energy diference tends to be slightly more realistic \cite{perdew2017}.

\begin{figure}[h]
    \centering
    {\includegraphics[width=0.5\textwidth]{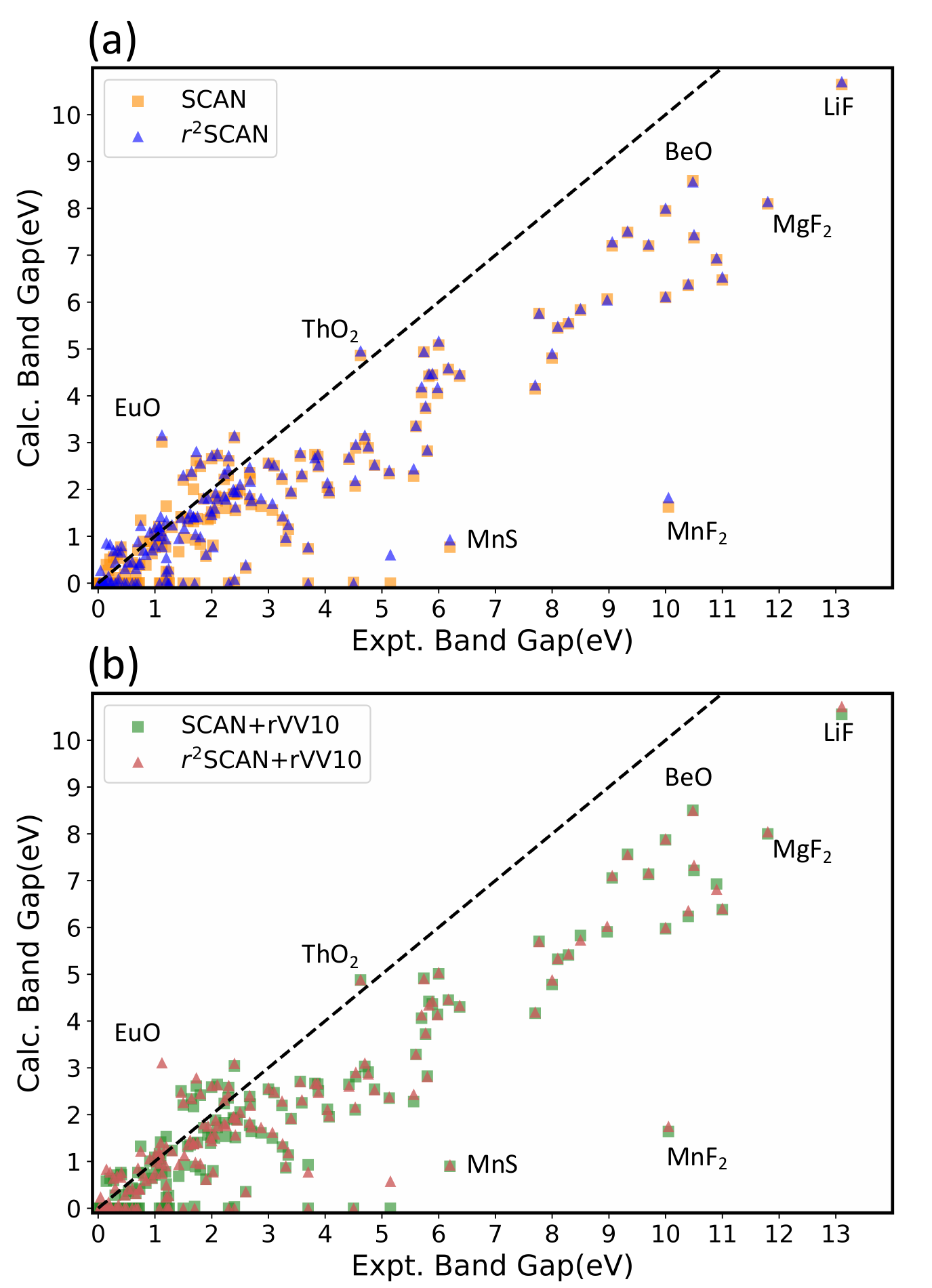}}
    \caption{Calculated and experimental electronic band gap.
    The dashed line corresponds to perfect agreement with experiment.
    Panel (a) plots SCAN and \rrscan, panel (b) plots SCAN+rVV10 and \rrscan{}+rVV10.}
\label{fig:bandgap}
\end{figure}

Figure \ref{fig:bandgap} compares computed band gaps to experimental values from Refs. \citenum{strehlow1973compilation,yang2016more}. Nearly all the points lie below the dashed line of perfect agreement, indicating that r$^2$SCAN systematically underestimates the fundamental gap.
However, consistent with Ref. \citenum{aschebrock2019}, some of the r$^2$SCAN gaps are larger than those predicted by SCAN: SCAN predicts WS$_2$ to be gapless, whereas \rrscan{} predicts a 1.41 eV gap, slightly larger than the 1.1 eV experimental gap.
Similar trends are seen for the compounds ZnTe, Sb$_2$Te$_3$, InSe, InSb, InN, InAs, GeTe, FeS$_2$, and GaAs. 

There are a few systems where \rrscan{} overestimates the gap more than SCAN underestimates it.
For example, SnSe has an experimental bandgap of 0.91 eV; SCAN predicts a 0.89 eV gap, whereas \rrscan predicts a much larger 1.06 eV gap. 
A similar tendency to overestimate the gaps of small-gap insulators has been observed \cite{neupane2021} for the TASK \cite{aschebrock2019} meta-GGA, which was designed for accurate bandgap prediction.
The \rrscan{} bandgap tends to be more accurate across all insulators than SCAN's: \rrscan{} (SCAN) makes a 1.15 (1.20) eV MAE for this set.
The 0.05 eV difference in average errors is largely due to wide-gap compounds such as LiF, MgF$_2$, BeO and MnF$_2$.
For insulators with an experimental gap less than 5 eV, \rrscan (SCAN) makes a 0.73 (0.77) eV MAE; for insulators with experimental gaps greater than 5 eV, \rrscan{} (SCAN) makes a 1.36 (1.43) eV MAE.

\section{Decomposition enthalpy}

Recent studies showed that the signs of decomposition enthalpies are more useful quantities than formation enthalpies for evaluating the stability of compounds\cite{bartel2019role,bartel2022review}. To calculate decomposition enthalpies, we must evaluate the reaction energies of the competing phases of compounds and elements in a composition space\cite{hautier2012accuracy, ong2008li,zunger2018inverse}. 
For a given ternary compound ABC, the compound ABC competes with all the possible elements, binaries, and ternaries in the corresponding A-B-C space. To obtain the decomposition enthalpy of ternary ABC, we compare the energy of ABC with the linear combination of the competing compounds with the same average composition as the ABC compound that minimizes the combined energy of the competing compounds, $E_\mathrm{A-B-C}$. The decomposition
enthalpy, $\Delta H_d$ is :

\begin{equation}
    \Delta H_{\mathrm{d}} = E_\mathrm{rxn} = E_\mathrm{ABC} - E_\mathrm{A-B-C}.
\end{equation}
$\Delta H_d > 0$ indicates that the ABC compound is unstable with respect to compounds formed from the competing space of A-B-C. Similarly, $\Delta H_d < 0$  indicates that the ABC compound is stable with respect to its competing phases.
 
The decomposition reactions that determine $\Delta H_d$ fall into one of three types as defined in Ref.\citenum{bartel2019role}. A type 1 compound is the only known compound in its composition space; the decomposition products are its elemental constituents, and thus $\Delta H_d$ = $\Delta H_f$.
For Type 2 compounds, the decomposition products are compounds, thus there are no elemental constituents in the decomposed products. For Type 3 compounds, the decomposition products are a combination of compounds and elements.

\begin{figure*}
    \centering
    {\includegraphics[width=0.83\textwidth]{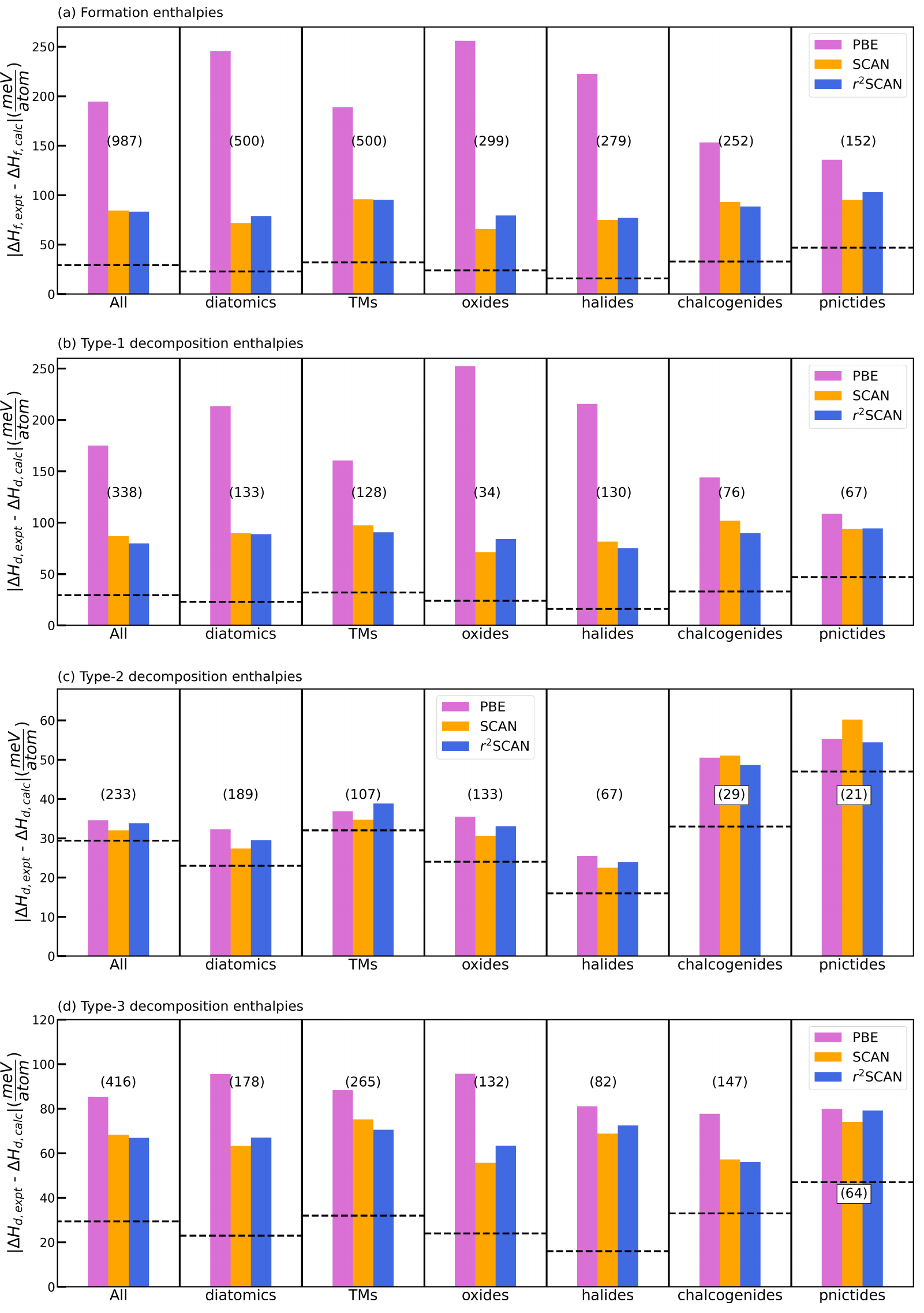}}
    \caption{ The mean absolute error for PBE, SCAN and r$^2$SCAN taken with respect to experimental values \cite{bartel2019role,bale2014recent} for (a) Formation enthalpies (b) Type 1 decomposition enthalpies (c) Type 2 decomposition enthalpies (d) Type 3 decomposition enthalpies.
    In addition to the full set (``All''), we consider the diatomic subset, where compounds contain at least one element in the set H, N, O, F, Cl; the TMs subset, compounds that contain at least one element from groups 3--11; oxides, compounds that contain oxygen; halides, compounds that contain at least one element in the set F, Cl, Br, I; chalcogenides, compounds that contain at least one element in the set S, Se, Te; pnictides, compounds that contain at least one element in the set N, P, As, Sb, Bi. The numbers in parentheses above each set of bars indicate the number of compounds in that subset. The dashed horizontal line indicates the approximate uncertainty of $\Delta H_\text{f,expt}$ or $\Delta H_\text{d,expt}$.}
    \label{fig:errorstatistics_fe_decomp}
\end{figure*}

Here we compare the performance of PBE, SCAN and \rrscan for the decomposition enthalpies of solid-state materials previously benchmarked by Bartel, \textit{et al.} \cite{bartel2019role}. To better elucidate the accuracy of the decomposition enthalpies, Figure \ref{fig:errorstatistics_fe_decomp} presents the errors for subsets of the database. ``All'' is the entire 987 solid set; ``diatomics'' contain at least one  element in the set H, N, O, F, Cl; ``TMs'' contain at least one element from groups 3-11; ``oxides'' contain oxygen; ``halides'' contain F, Cl, Br, or I; ``chalcogenides'' contain S, Se, or Te; and ``pnictides'' contain N, P, As, Sb, or Bi. The total set of 987 solids is partitioned into Type 1 (34\%), 2 (24\%), and 3 (42\%) reactions. As shown in Fig. \ref{fig:errorstatistics_fe_decomp}(a), we first analyzed $\Delta H_f$ for all compounds to establish a baseline for subsequent comparison to  $\Delta H_d$. The MAE for $\Delta H_f$ is partitioned for various chemical subsets of the dataset in Fig \ref{fig:errorstatistics_fe_decomp}(a) to understand elemental dependence. For this set of 987 compounds, the MAE between the experimentally determined $\Delta H_f$ at 298 K, and calculated $\Delta H_f$ at 0 K, was found to be 194 meV/atom for PBE, 84 meV/atom for SCAN, and 83 meV/atom for r$^2$SCAN. PBE shows large systematic errors for a range of diversely bonded systems. SCAN and \rrscan are comparably accurate for all the partitioned subsets, except for oxides, which are described better by SCAN. The good general accuracy of SCAN is typically attributed to its satisfaction of all 17 known constraints applicable to a semilocal DFA \cite{kaplan2023}.
\rrscan{} satisfies one fewer exact constraint than SCAN by recovering a lower-order gradient expansion for exchange than SCAN \cite{furness2020}. \rrscan{}'s smoother exchange-correlation energy density could be the reason for its exceptional performance.

To determine the decomposition enthalpies $\Delta H_d$, and thus the thermodynamic stability of compounds, we used $\Delta H_f$ to perform an $N$-dimensional convex hull analysis.
We consider only PBE, SCAN, \rrscan{}, and experimental values.
For 338 compounds that decompose as Type 1 reactions, $\Delta H_f$ = $\Delta H_d$, the 80 meV/atom MAE of \rrscan is the lowest of all considered DFAs, followed by 87 meV/atom for SCAN, and 175 meV/atom for PBE. As expected, the trend for the Type 1 reactions is  similar to the overall formation enthalpies shown in figure \ref{fig:errorstatistics_fe_decomp}(a).  In ``real'' phase diagrams that are comprised only of computed data (e.g., those retrievable in Materials Project\cite{jain2013commentary}, OQMD\cite{saal2013materials,kirklin2015open}, etc.), there are effectively zero “Type 1” compounds because (nearly) every chemical space has at least two calculated compositions.

For the 233 Type 2 decomposition reactions, where compounds compete only with other compounds and not elements, r$^2$SCAN, SCAN and PBE are found to perform comparably, with MAEs of $\sim$35 meV/atom.
All DFAs have slightly larger MAEs for the Type 2 chalcogenide and pnictide decomposition enthalpies.  Specifically for Type 2, our results show excellent agreement between experiment and theory for $\Delta H_d$ on a diverse set of materials without requiring an empirical Hubbard-like $U$ correction. For the 416 Type 3 decomposition reactions, where compounds have elements and compounds that compete energetically, $\Delta H_d$ does not significantly change from SCAN to \rrscan. However, for these compounds, SCAN and \rrscan improve over PBE by $\sim$20\%, and the MAE between \rrscan and experiment (67 meV/atom) falls between those for Type 1 (79 meV/atom) and Type 2 (34 meV/atom).

\section{Conclusions}

This work has shown that \rrscan{} \cite{furness2022} and the dispersion-corrected \rrscan{}+rVV10 \cite{ning2022} are suitable for general-purpose solid-state materials discovery, in the vein of Refs. \citenum{isaacs2018}, \citenum{zhang2018}, and \citenum{kingsbury2022}.
Thus we highlight conclusions common to previous works and ours, and those that are unique to the work at hand.

Zhang, \textit{et al.} \cite{zhang2018} established that SCAN predicted formation enthalpies of 102 main group compounds with roughly a factor of 2.5 less average absolute error than PBE.
This greater-than-twofold decrease in MAE is readily confirmed with the larger 260-solid set of main group compounds presented in Fig. \ref{fig:MAD_fe}. Likewise, the formation enthalpies (or Type I decomposition enthalpies) of Figs. \ref{fig:errorstatistics_fe_decomp}(a-b) show a roughly 1.5 to 3-fold decrease in MAE in going from PBE to SCAN or \rrscan{}.

Figure \ref{fig:MAD_fe} and Table \ref{tab:magnetism} show that the simple PBE GGA is more accurate than the more sophisticated meta-GGAs for the formation enthalpies and magnetic moments of metals, as observed in earlier works \cite{isaacs2018,mejia2019analysis,ekholm2018,fu2018,kaplan2022}.
The reason has been discussed in Ref. \citenum{kaplan2022}: The exact exchange-correlation energy density at a position is proportional to the Coulomb interaction between an electron at that position and the density of the exact exchange-correlation hole which surrounds it.
The more short-ranged the hole shape is, the better the functional can be approximated using just the local electron density and its low-order derivatives. 
Since the long-range part of the exact exchange hole is screened by the long-wavelength dielectric constant of the material, the hole shape is especially short-ranged in metals, where this screening is perfect, and where the meta-GGA ingredient $\tau$ is somewhat too nonlocal. 
Global hybrid functionals, with the even-more-nonlocal exact exchange energy density as an ingredient, are even less accurate \cite{fu2018} for the magnetic moments of metals than meta-GGAs are.

Recall that SCAN recovers more of the intermediate-ranged vdW interaction than does \rrscan{}.
A larger rVV10 $b$-parameter (see also Refs. \citenum{peng2016} and \citenum{ning2022}) more strongly damps the dispersion correction at short-range.
The SCAN+rVV10 value $b=15.7$ \cite{peng2016} is much larger than that of \rrscan{}+rVV10, $b=11.95$, \cite{ning2022} indicating that \rrscan{} needs a more substantial dispersion correction at short- to intermediate-range than does SCAN.
In this sense, rVV10 is a more compatible correction to \rrscan{} than to SCAN (rVV10 essentially over-corrects SCAN at shorter range).
Thus, while Fig. \ref{fig:MAD_fe} often shows marked increases in the MAEs for SCAN+rVV10 over SCAN (see especially the strongly-bound, weakly-bound, and oxide MAEs), \rrscan{}+rVV10 essentially does no harm to \rrscan{} in predicting formation enthalpies.
Moreover, \rrscan{}+rVV10 predicts modestly more accurate cell volumes than \rrscan{}, as shown in Table \ref{tab:veq_errs}.

We found that both \rrscan{} and SCAN tend to underestimate the fundamental bandgaps of insulators, as noted previously \cite{kingsbury2022,Mejia-Rodriguez2020,kaplan2022}.
However, we also found that \rrscan{} sometimes overestimates the bandgaps of narrow-gap insulators.
This is consistent with the tendency of the TASK meta-GGA \cite{aschebrock2019} to overestimate the gaps of narrow-gap insulators \cite{neupane2021}.

We confirm the conclusion of Ref. \citenum{kingsbury2022} that \rrscan{} predicts slightly more accurate formation enthalpies and cell volumes than SCAN.
GGAs tend to predict much more accurate formation enthalpies for weakly-bound solids, as shown here, and in Ref. \citenum{isaacs2018} for PBE and SCAN, and in Ref. \citenum{kingsbury2022} for PBEsol \cite{perdew2008}, SCAN, and \rrscan{}.
Likewise, PBE and PBEsol predict much more accurate energetics of transition metal intermetallics \cite{isaacs2018,kingsbury2022,kaplan2022} than the meta-GGAs, for reasons discussed previously.

The Type 2 decomposition enthalpies of Fig.\ref{fig:errorstatistics_fe_decomp} show PBE is slightly more accurate than SCAN or \rrscan.

However, all DFAs predict Type 2 decomposition enthalpies with accuracy close to or below the 30 meV/atom experimental uncertainty. Except for the diatomic and oxide decomposition enthalpies, much smaller decreases in the Type 3 decomposition enthalpies are observed in going from the GGA to meta-GGA level.
We have not applied SCAN+rVV10 and \rrscan{}+rVV10 to the set of solid-state decomposition enthalpies (which differs from the set \cite{isaacs2018} presented in Fig. \ref{fig:MAD_fe}), for two reasons: (1) this would be computationally cost-prohibitive; and (2) if most solids in the set are strongly-bound, a dispersion correction will make insignificant changes to the total energies.

Given the general accuracy and numerical stability of \rrscan{} \cite{furness2022,Mejia-Rodriguez2020,kingsbury2022} and \rrscan{}+rVV10 \cite{ning2022}, it is safe to recommend either for general materials discovery.
When transition metal intermetallic systems are of interest, a GGA or Laplacian-level meta-GGA may be a better choice.
When considering layered materials, we recommend \rrscan{}+rVV10.

\begin{acknowledgments}

J.S., J.N., and M.K. acknowledge the support of the U.S. Department of Energy (DOE), Office of Science (OS), Basic Energy Sciences (BES), Grant No. DE-SC0014208.
J.W.F. acknowledges support from DOE grant DE-SC0019350.
A.D.K. thanks Temple University for a Presidential Fellowship.
J.P.P. acknowledges the support of the US NSF under Grant No. DMR-1939528.

We thank Dr. Alan Weimer for providing experimental formation enthalpy data.

\end{acknowledgments}
\bibliography{r2SCAN}
\bibliographystyle{apsrev4-1}

\clearpage

\onecolumngrid

\renewcommand{\thepage}{S\arabic{page}}
\renewcommand{\thesection}{S\arabic{section}}
\renewcommand{\theequation}{S\arabic{equation}}
\renewcommand{\thetable}{S\arabic{table}}
\renewcommand{\thefigure}{S\arabic{figure}}
\renewcommand{\epsilon}{\varepsilon}

\clearpage
\setcounter{section}{0}
\setcounter{page}{0}
\setcounter{equation}{0}
\setcounter{table}{0}
\setcounter{figure}{0}

\date{\today}
\widetext
\begin{flushleft}

\textbf{\Large Supplemental Materials: Testing the r$^2$SCAN  density functional for the
thermodynamical stability of solids with and without a van der
Waals correction}\\

\end{flushleft}

\begin{flushleft}
{Manish Kothakonda,$^{1}$  Aaron Kaplan,$^{2}$,Eric B. Isaacs,$^{3}$Christopher J Bartel,$^{4}$ James Furness,$^{1}$ Jinliang Ning,$^{1}$ Chris Wolverton,$^{5}$John Perdew,$^{2}$ Jianwei Sun$^{1}$\\

\textit{\normalsize{$^{1}$Department of Physics and Engineering Physics, Tulane University, New Orleans, LA 70118, USA}}\\
\textit{\normalsize{$^{2}$Department of Physics, Temple University, Philadelphia, Pennsylvania 19122, United States}}\\
\textit{\normalsize{$^{3}$HRL Laboratories, LLC, Malibu, California 90265, United States}}\newline
\textit{\normalsize{$^{4}$Department of Materials Science and Engineering, University of California, Berkeley, United States}}\newline
\textit{\normalsize{$^{5}$Department of Materials Science and Engineering, Northwestern University, Evanston, Illinois 60208, United States}}\newline

\normalsize{ 

E-mail:  jsun@tulane.edu}

}
\end{flushleft}

\clearpage

\begin{figure}[htp]
    \centering
    {\includegraphics[width=0.48\textwidth]{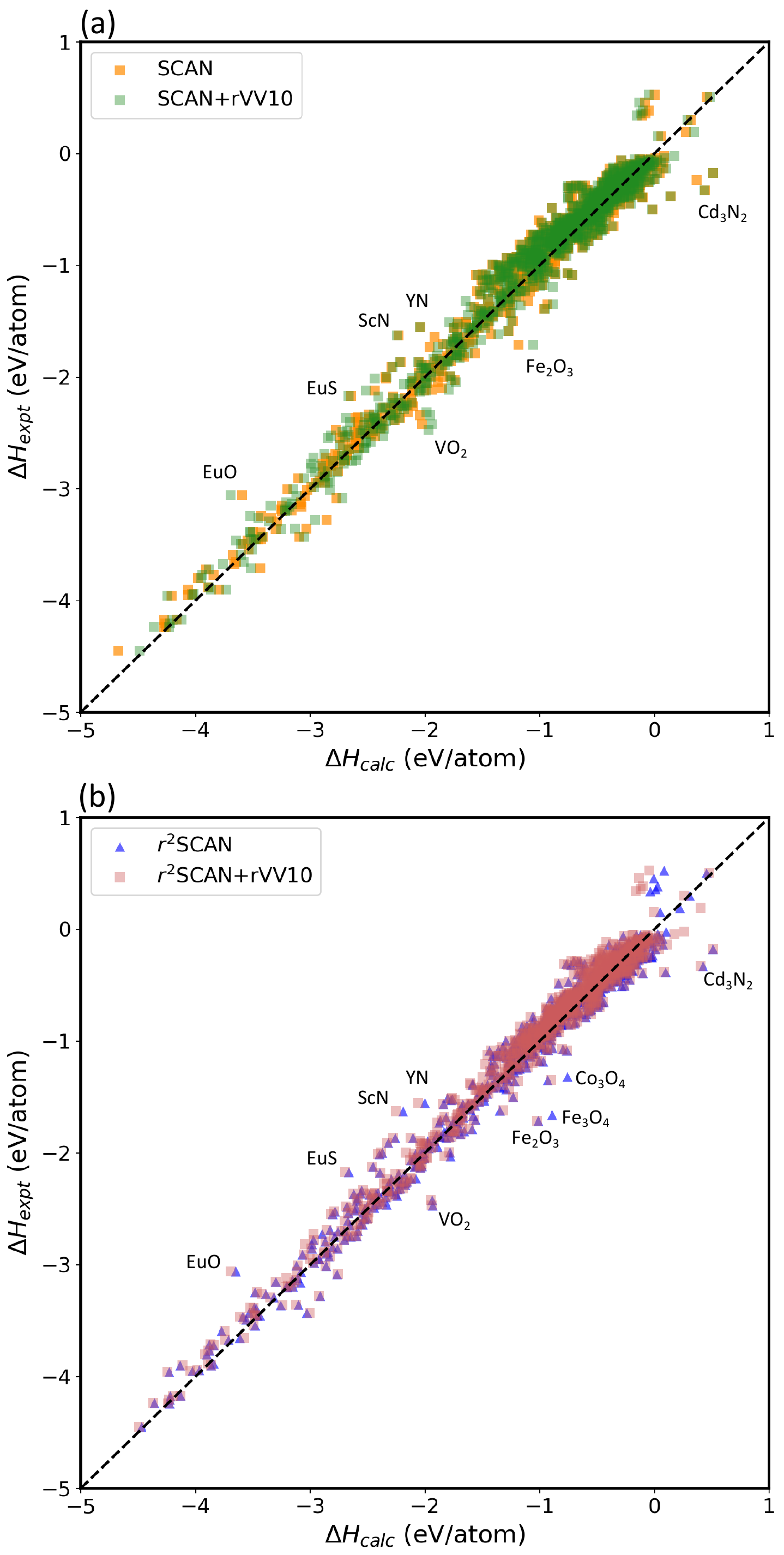}}
    \caption{Comparison of calculated and experimental formation enthalpies for the 1015 compounds for (a) SCAN and SCAN+rVV10 (b) \rrscan and \rrscan+rVV10. Multiple points for the same compound and functional correspond to different sources of experimental formation enthalpy. The dashed diagonal line corresponds to the $\Delta H_\mathrm{calc} = \Delta H_\mathrm{expt}$ line of perfect agreement. }
    \label{fig:sm_fe}
\end{figure}

\begin{figure*}[htp]
    \begin{center}
    \centering
    {\includegraphics[width=0.55\textwidth]{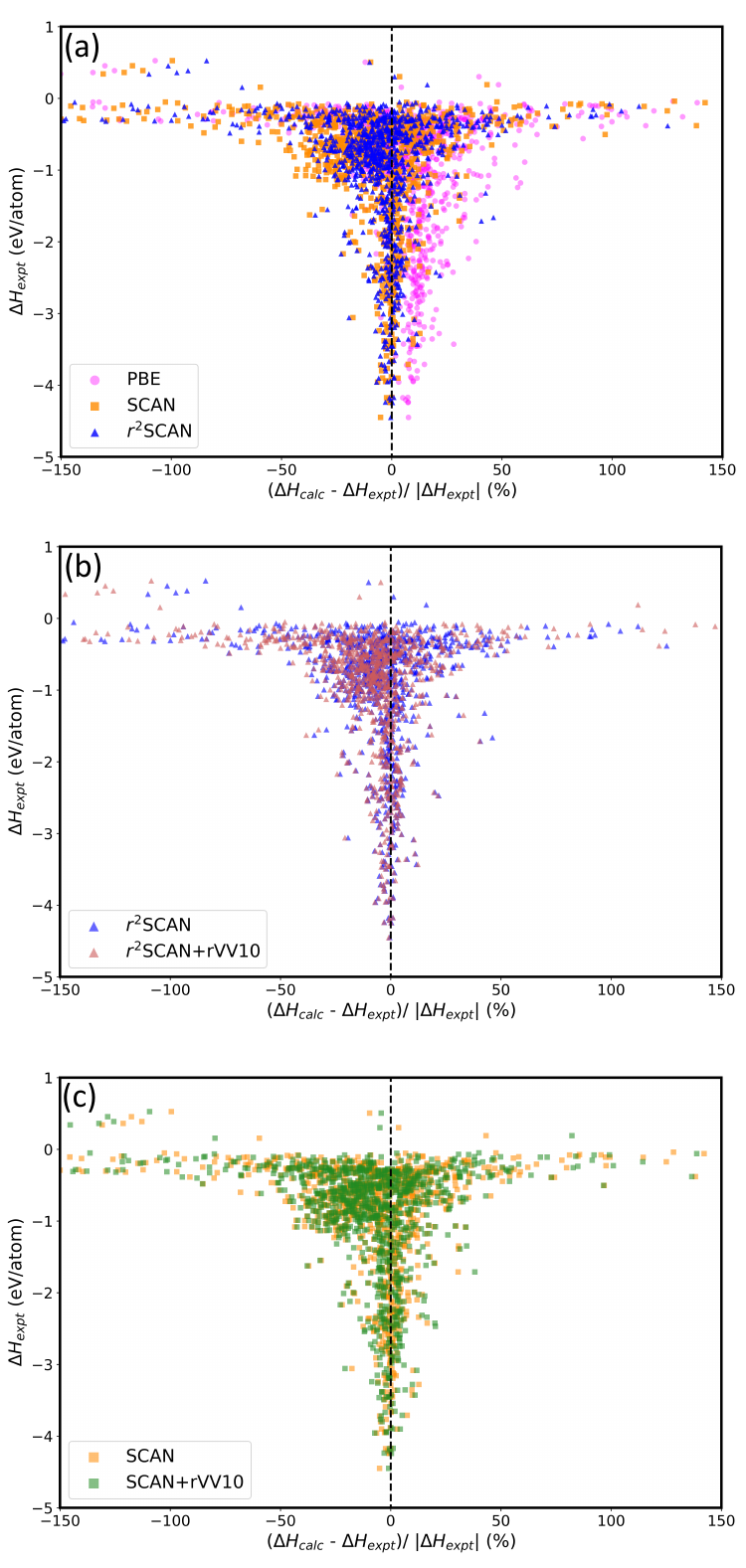}}
    \end{center}
    \caption{Relative error of the calculated formation enthalpy plotted against the experimental formation enthalpy. The dashed vertical lines correspond to the $\Delta H_\mathrm{calc}$ = $\Delta H_\mathrm{expt}$ line of perfect agreement. For the relative errors, the range is limited to $\pm 150\%$.}
    \label{fig:abs_rel_error}
\end{figure*}

\begin{figure*}[htp]
    \begin{center}
    \centering
    {\includegraphics[width=0.8\textwidth]{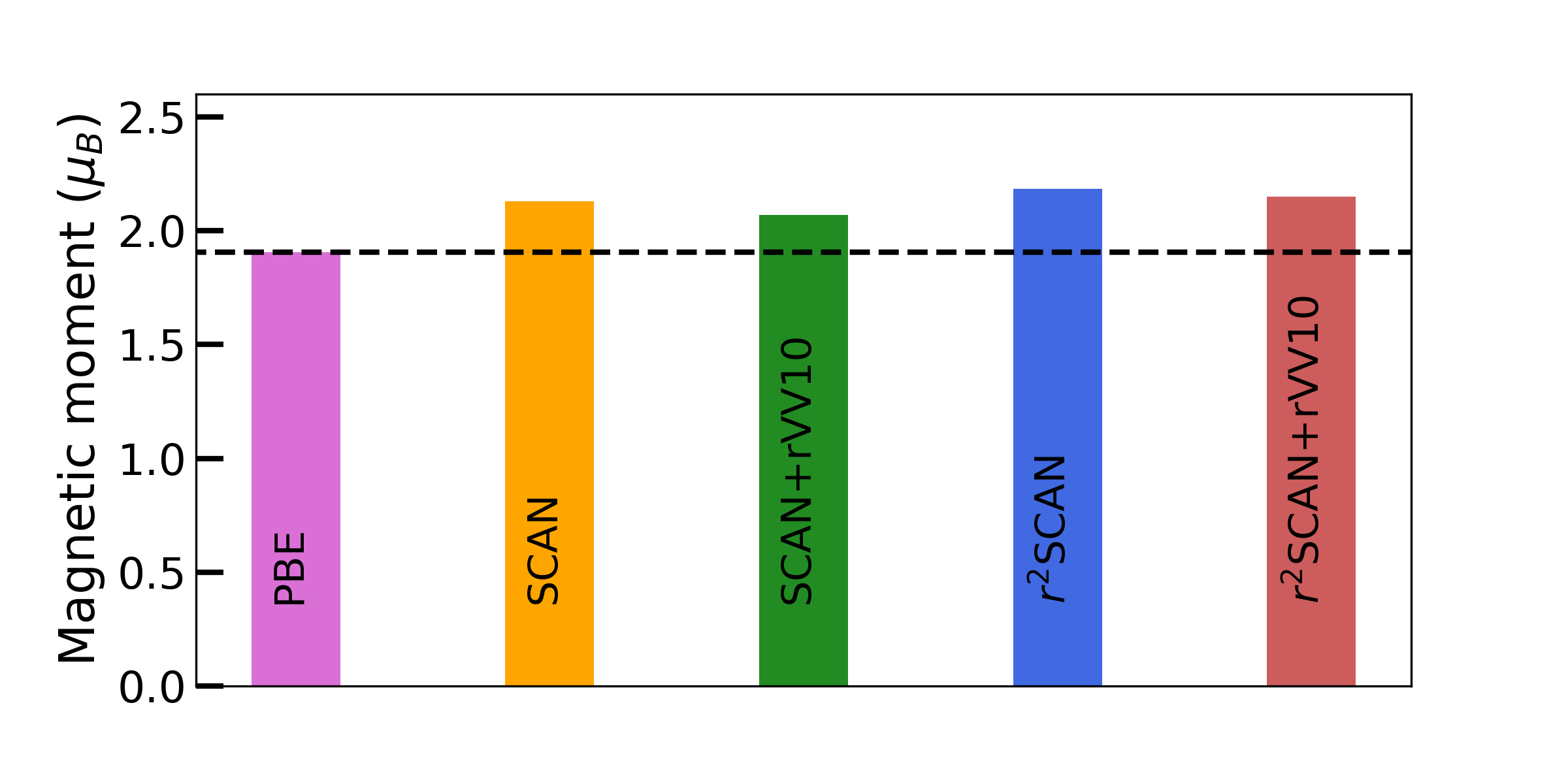}}
    \end{center}
    \caption{Average Magnetic moment of 157 magnetic compounds, the horizontal line is leveled with PBE magnetic moment to notice the difference.}
    \label{fig:avgmagentimoment}
\end{figure*}

\clearpage

\begin{longtable}{|p{0.5in} |p{0.5in} |p{0.5in}| p{0.5in} |p{1in}| p{1in}|}
\caption{Comparison of approximate bandgaps (eV) to experiemnt.}

\\

\hline \multicolumn{1}{|c|}{\textbf{System}} & \multicolumn{1}{c|}{\textbf{Expt}} & \multicolumn{1}{c|}{\textbf{SCAN}} & \multicolumn{1}{c|}{\textbf{\rrscan{}}} & \multicolumn{1}{c|}{\textbf{SCAN+rVV10}} & \multicolumn{1}{c|}{\textbf{\rrscan{}+rVV10}}
\\ \hline 
\endfirsthead

\multicolumn{6}{c}%
{{\bfseries \tablename\ \thetable{} -- continued from previous page}} \\
\hline \multicolumn{1}{|c|}{\textbf{System}} & \multicolumn{1}{c|}{\textbf{Expt}} & \multicolumn{1}{c|}{\textbf{SCAN}} & \multicolumn{1}{c|}{\textbf{\rrscan{}}} & \multicolumn{1}{c|}{\textbf{SCAN+rVV10}} & \multicolumn{1}{c|}{\textbf{\rrscan{}+rVV10}}\\ \hline 
\endhead

\hline \multicolumn{6}{|r|}{{Continued on next page}} \\ \hline
\endfoot

\hline \hline
\endlastfoot

\hline
        
Ag2O   & 1.2    & 0.245 & 0.261 & 0.228 & 0.244 \\

Ag2S   & 0.95   & 0.63  & 0.713 & 0.64  & 0.651 \\

AgCl   & 3.25   & 1.346 & 1.437 & 1.309 & 1.383 \\

AgI    & 2.869 & 1.641 & 1.808 & 1.61  & 1.727 \\

Al2O3  & 9.7    & 7.196 & 7.237 & 7.137 & 7.17  \\
 
Al2Se3 & 3.1    & 2.502 & 2.526 & 2.469 & 2.489 \\

AlAs   & 2.1    & 1.754 & 1.812 & 1.728 & 1.762 \\

AlN    & 5.74   & 4.933 & 4.946 & 4.915 & 4.906 \\
 
AlP    & 2.45   & 1.911 & 1.952 & 1.888 & 1.888 \\
 
AlSb   & 1.62   & 1.382 & 1.497 & 1.352 & 1.456 \\
        
As     & 1.2    & 0     & 0.546 & 0.514 & 0.497 \\
        
AsI3   & 2.29   & 2.297 & 2.441 & 2.356 & 2.42  \\
        
BaF2   & 9.06   & 7.203 & 7.288 & 7.057 & 7.109 \\
        
BaO    & 5.13   & 2.336 & 2.41  & 2.353 & 2.38  \\
        
BaS    & 3.88   & 2.489 & 2.525 & 2.508 & 2.478 \\
        
BAs    & 1.46   & 1.415 & 1.404 & 2.508 & 2.478 \\
        
BaTe   & 3.4    & 1.915 & 1.971 & 1.911 & 1.926 \\
        
BeO    & 10.48  & 8.591 & 8.57  & 8.506 & 8.499 \\
        
Bi     & 0.015  & 0     & 0     & 0     & 0     \\
        
Bi2Se3 & 0.21   & 0.51  & 0.831 & 0.637 & 0.778 \\
        
Bi2Te3                       & 0.145    & 0.392    & 0.857      & 0.579          & 0.836            \\
        
BiI3   & 1.73   & 2.615 & 2.814 & 2.61  & 2.787 \\
        
BN     & 8      & 4.804 & 4.905 & 4.78  & 4.882 \\
        
BP     & 2      & 1.52  & 1.467 & 1.525 & 1.446 \\
        
CaB6   & 4.5    & 0.014 & 0     & 0     & 0     \\
        
CaF2   & 10     & 7.944 & 8.006 & 7.87  & 7.896 \\
        
CaI2   & 5.98   & 4.051 & 4.176 & 4.143 & 4.136 \\
        
CaO    & 7.7    & 4.148 & 4.231 & 4.166 & 4.193 \\
        
CaS    & 5.8    & 2.816 & 2.847 & 2.824 & 2.816 \\
        
CaSe   & 4.87   & 2.514 & 2.535 & 2.541 & 2.534 \\
        
CaTe   & 4.07   & 1.926 & 1.979 & 1.954 & 1.978 \\
        
CdCl2  & 5.7    & 4.069 & 4.196 & 4.06  & 4.135 \\
        
CdO    & 1.2    & 0     & 0.075 & 0     & 0     \\
        
CdS    & 2.4175 & 1.554 & 1.631 & 1.51  & 1.568 \\
        
CdSe   & 1.714  & 0.925 & 1.044 & 0.891 & 0.981 \\
        
CdTe   & 1.517  & 0.967 & 1.171 & 0.925 & 1.128 \\
        
CeN    & 0.7    & 0     & 0     & 0     & 0     \\
        
CeO2   & 2.68   & 2.222 & 2.179 & 2.224 & 2.2   \\
        
CoO    & 0.47   & 0.192 & 0.313 & 0.287 & 0.287 \\
        
CoSi   & 0.045  & 0     & 0     & 0     & 0     \\
        
Cr2O3  & 1.68   & 2.004 & 1.418 & 2.172 & 1.384 \\
        
CrO2   & 0.23   & 0     & 0     & 0     & 0     \\
        
CrSi2  & 0.35   & 0     & 0     & 0     & 0     \\
        
CsCl   & 8.1    & 5.447 & 5.485 & 5.319 & 5.342 \\
        
CsF    & 10     & 6.1   & 6.122 & 5.972 & 6.002 \\
        
CsI    & 6.37   & 4.42  & 4.472 & 4.302 & 4.338 \\
        
Cu2O   & 2.023  & 0.81  & 0.781 & 0.799 & 0.78  \\
        
Cu2Se  & 1.23   & 0     & 0     & 0     & 0     \\
        
Cu2Te  & 1.08   & 0     & 0     & 0     & 0     \\
        
CuCl   & 3.306  & 0.899 & 0.98  & 0.863 & 0.901 \\
        
CuI    & 3.07   & 1.559 & 1.707 & 1.497 & 1.623 \\
        
EuO    & 1.122  & 3.011 & 3.164 & 0     & 3.111 \\
        
EuS    & 1.645  & 2.309 & 2.386 & 2.339 & 2.348 \\
        
FeI2   & 5.15   & 0     & 0.607 & 0     & 0.578 \\
        
FeP2   & 0.4    & 0.769 & 0.666 & 0.762 & 0.666 \\
        
FeS2   & 1.2    & 1.641 & 1.333 & 1.535 & 1.317 \\
        
FeSi   & 0.1    & 0     & 0     & 0     & 0     \\
        
FeTe2  & 0.46   & 0     & 0     & 0     & 0     \\
        
Ga2O3  & 4.54   & 2.884 & 2.964 & 2.812 & 2.906 \\
        
Ga2S3  & 3.59   & 2.275 & 2.337 & 2.253 & 2.318 \\
        
Ga2Se3 & 2.05   & 1.503 & 1.605 & 1.498 & 1.589 \\
        
GaAs   & 1.42   & 0.669 & 0.957 & 0.682 & 0.937 \\
        
GaN    & 3.24   & 2.221 & 2.323 & 2.195 & 2.291 \\
        
GaP    & 2.22   & 1.824 & 1.864 & 1.823 & 1.809 \\
        
GaS    & 2.5    & 1.985 & 2.108 & 2.053 & 2.057 \\
        
GaSb   & 0.725  & 0.008 & 0.405 & 0     & 0.406 \\
        
GaSe   & 1.98   & 1.388 & 1.543 & 1.391 & 1.517 \\
        
Ge     & 0.665  & 0.138 & 0.313 & 0.313 & 0.313 \\
        
GeI2   & 1.5    & 2.198 & 2.306 & 2.204 & 2.261 \\
        
GeO2   & 5.56   & 2.283 & 2.444 & 2.278 & 2.436 \\
        
GeS    & 1.58   & 1.321 & 1.379 & 1.322 & 1.361 \\
        
GeSe   & 1.1    & 1.005 & 1.122 & 1.048 & 1.085 \\
        
GeSe2  & 2.38   & 1.897 & 2     & 1.934 & 1.958 \\
        
GeTe   & 0.84   & 0.387 & 0.619 & 0.534 & 0.597 \\
        
I      & 1.3    & 1.194 & 1.248 & 1.233 & 1.233 \\
        
InAs   & 0.356  & 0     & 0.094 & 0     & 0.078 \\
        
InN    & 2.4    & 0.015 & 0.086 & 0.028 & 0     \\
        
InS    & 1.86   & 1.797 & 1.821 & 1.721 & 1.805 \\
        
InSb   & 0.17   & 0     & 0.044 & 0     & 0     \\
        
InSe   & 1.187  & 0.775 & 0.948 & 0.778 & 0.912 \\
        
K2S    & 2.1    & 2.72  & 2.774 & 2.646 & 2.64  \\
        
K2Se   & 1.8    & 2.494 & 2.565 & 2.416 & 2.454 \\
        
K3Sb   & 1      & 0.761 & 0.817 & 0.712 & 0.752 \\
        
KCl    & 8.5    & 5.83  & 5.864 & 5.827 & 5.735 \\
        
KF     & 10.9   & 6.899 & 6.946 & 6.928 & 6.819 \\
        
KI     & 6.17   & 4.559 & 4.601 & 4.448 & 4.469 \\
        
Li3Sb  & 1      & 1.056 & 1.185 & 1.088 & 1.151 \\
        
LiCl   & 9.33   & 7.483 & 7.519 & 7.565 & 7.558 \\
        
LiF                          & 13.105   & 10.643   & 10.705     & 10.554         & 10.722           \\
        
LiI    & 6      & 5.083 & 5.169 & 5.01  & 5.038 \\
        
Mg2Ge  & 0.532  & 0.359 & 0.454 & 0.355 & 0.415 \\
        
Mg2Pb  & 0.041  & 0     & 0.271 & 0     & 0.24  \\
        
Mg2Si  & 0.6    & 0.429 & 0.456 & 0.429 & 0.42  \\
        
MgF2   & 11.8   & 8.096 & 8.147 & 7.998 & 8.044 \\
        
MgO    & 7.77   & 5.753 & 5.761 & 5.702 & 5.701 \\
        
MgSe   & 5.6    & 3.345 & 3.366 & 3.287 & 3.296 \\
        
MgTe   & 4.7    & 3.079 & 3.162 & 3.03  & 3.103 \\
        
MnF2   & 10.05  & 1.623 & 1.828 & 1.639 & 1.751 \\
        
MnI2   & 4.04   & 2.048 & 2.146 & 2.114 & 2.109 \\
        
MnO    & 3.7    & 0     & 0     & 0     & 0     \\
        
MnS    & 6.2    & 0.768 & 0.928 & 0.894 & 0.927 \\
        
MnSe   & 1.8    & 0.834 & 0.992 & 0.816 & 0.956 \\
        
MnTe   & 1.25   & 0     & 0     & 0     & 0     \\
        
MoS2   & 1.07   & 1.207 & 1.216 & 1.077 & 1.216 \\
        
Na2S   & 2.4    & 3.103 & 3.157 & 3.039 & 3.098 \\
        
Na2Se  & 2      & 2.656 & 2.727 & 2.588 & 2.623 \\
        
Na2Te  & 2.3    & 2.612 & 2.724 & 2.579 & 2.625 \\
        
Na3Sb  & 1.1    & 0.949 & 1.019 & 0.91  & 0.963 \\
        
NaCl   & 8.97   & 6.065 & 6.047 & 5.906 & 6.027 \\
        
NaF    & 10.5   & 7.371 & 7.442 & 7.22  & 7.329 \\
        
NaI    & 5.89   & 4.443 & 4.479 & 4.368 & 4.425 \\
        
NiO    & 3.7    & 0.731 & 0.779 & 0.927 & 0.777 \\
        
NiS    & 0.12   & 0     & 0     & 0     & 0     \\
        
PbO    & 1.936  & 1.359 & 1.813 & 1.567 & 1.757 \\
        
PbO2   & 1.7    & 0     & 0     & 0.037 & 0     \\
        
PbS    & 0.41   & 0.619 & 0.802 & 0.716 & 0.767 \\
        
PbSe   & 0.27   & 0.555 & 0.695 & 0.594 & 0.644 \\
        
PdO    & 1.5    & 0     & 0     & 0     & 0     \\
        
PrO2   & 0.66   & 0     & 0     & 0     & 0     \\
        
PtS    & 0.8    & 0.888 & 0.71  & 0.752 & 0.71  \\
        
PtS2   & 0.75   & 1.343 & 1.237 & 1.323 & 1.221 \\
        
RbCl   & 8.29   & 5.538 & 5.579 & 5.411 & 5.436 \\
        
RbF    & 10.4   & 6.36  & 6.392 & 6.235 & 6.358 \\
        
RbI    & 5.83   & 4.415 & 4.471 & 4.422 & 4.348 \\
        
ReSi2  & 0.12   & 0     & 0     & 0     & 0     \\
        
S      & 3.82   & 2.745 & 2.681 & 2.669 & 2.669 \\
        
Sb     & 0.1    & 0     & 0     & 0     & 0     \\
        
Sb2Te3 & 0.3    & 0.11  & 0.449 & 0.277 & 0.439 \\
        
SbI3   & 2.22   & 2.215 & 2.346 & 2.237 & 2.317 \\
        
ScN    & 2.6    & 0.322 & 0.394 & 0.354 & 0.354 \\
        
Se     & 1.75   & 1.372 & 1.425 & 1.408 & 1.407 \\
        
Si     & 1.12   & 0.827 & 0.787 & 0.767 & 0.748 \\
        
SiO2   & 11     & 6.473 & 6.538 & 6.379 & 6.411 \\
        
SmS    & 0.22   & 0     & 0     & 0     & 0     \\
        
SnI2   & 2.4    & 1.903 & 1.964 & 1.902 & 1.941 \\
        
SnO2   & 2.7    & 1.679 & 1.767 & 1.642 & 1.742 \\
        
SnS    & 1.08   & 1.075 & 1.206 & 1.167 & 1.173 \\
        
SnS2   & 2.07   & 1.855 & 1.932 & 1.881 & 1.889 \\
        
SnSe   & 0.91   & 0.887 & 1.095 & 1.006 & 1.062 \\
        
SnSe2  & 1.03   & 0.933 & 1.014 & 0.95  & 0.992 \\
        
SnTe   & 0.18   & 0.271 & 0.149 & 0     & 0.13  \\
        
SrO    & 5.77   & 3.728 & 3.785 & 3.719 & 3.747 \\
        
SrS    & 4.76   & 2.888 & 2.93  & 2.91  & 2.873 \\
        
SrSe   & 4.42   & 2.641 & 2.693 & 2.646 & 2.609 \\
        
TaN    & 2.3    & 0     & 0     & 0     & 0     \\
        
TaS2   & 0.1    & 0     & 0     & 0     & 0     \\
        
TbO2   & 0.5    & 0     & 0     & 0     & 0     \\
        
Te     & 0.332  & 0.531 & 0.71  & 0.709 & 0.696 \\
        
ThO2   & 4.625  & 4.859 & 4.961 & 4.876 & 4.886 \\
        
Ti2O3  & 0.02   & 0     & 0     & 0     & 0     \\
        
TiO2   & 3      & 2.556 & 2.572 & 2.541 & 2.572 \\
        
TiS2   & 1.24   & 0.23  & 0.31  & 0.272 & 0.271 \\
        
Tl2Te3 & 0.7    & 0.706 & 0.89  & 0.762 & 0.86  \\
        
TlCl   & 3.56   & 2.712 & 2.788 & 2.707 & 2.709 \\
        
TlI    & 2.67   & 2.355 & 2.477 & 2.386 & 2.409 \\
        
TlSe   & 0.73   & 0.391 & 0.446 & 0.41  & 0.422 \\
        
V2O3   & 0.1    & 0     & 0     & 0     & 0     \\
        
VO     & 0.3    & 0     & 0     & 0     & 0     \\
        
WS2    & 1.1    & 0     & 1.423 & 1.411 & 1.407 \\
        
YN     & 1.9    & 0.576 & 0.62  & 0.621 & 0.62  \\
        
ZnI2   & 4.53   & 2.071 & 2.194 & 2.102 & 2.162 \\
        
ZnO    & 3.35   & 1.151 & 1.252 & 1.138 & 1.195 \\
        
ZnS    & 3.87   & 2.704 & 2.744 & 2.653 & 2.672 \\
        
ZnSe   & 2.67   & 1.8   & 1.893 & 1.771 & 1.84  \\
        
ZnTe   & 2.25   & 1.598 & 1.797 & 1.534 & 1.774 \\
        
ZrC    & 0.6    & 0     & 0     & 0     & 0     \\
        
ZrS2   & 1.68   & 1.314 & 1.417 & 1.388 & 1.383 \\\hline 
\end{longtable}


\clearpage

\end{document}